\documentstyle[12pt]{article}
\def\ZZ{\hbox{\it Z\hskip -4.pt Z}}
\def\RR{\hbox{\it I\hskip -2.pt R }}
\newcommand{\nc}{\newcommand}
\nc{\beq}{\begin{equation}}
\nc{\eeq}{\end{equation}}
\nc{\beqa}{\begin{eqnarray}}
\nc{\eeqa}{\end{eqnarray}}
\nc{\noi}{\noindent}

\begin{document}

\title{
\begin{flushright}
{$\,$}\\[-3cm]
{\small USACH-FM-00-02}\\[2cm]
\end{flushright}
{\bf Bosonized 
supersymmetric quantum mechanics 
and supersymmetry
of parabosons 
(parafermions)\thanks{\it Talk given at the International Workshop
``Supersymmetry and Quantum Symmetries", JINR, Dubna, Russia, July 26-31,1999}}}

\author{
{\sf Mikhail S. Plyushchay}\\
{\small {\it Departamento de F\'{\i}sica, 
Universidad de Santiago de Chile,}}\\
{\small {\it Casilla 307, Santiago 2, Chile}}\\
{\small {and}}\\
{\small {\it Institute for High Energy Physics, Protvino,
Russia}}\\
{\small {\it E-mail: mplyushc@lauca.usach.cl}}}
\date{}

\maketitle
\vskip-2.0cm

\begin{abstract}
We review the construction of minimally bosonized 
supersymmetric quantum mechanics
and its relation to hidden supersymmetries in pure
parabosonic (parafermionic) systems. 
\end{abstract}

The simplest $N=1$ supersymmetric quantum mechanical 
system is the superoscillator \cite{nic} given by the 
Hamiltonian
\beq
\label{hosc}
H=\frac{1}{2}\{b^+,b^-\}+\frac{1}{2}[f^+,f^-].
\eeq
In addition to $H$,
the system is characterized by 
the two conserved integrals of motion (supercharges)
\beq
\label{qosc}
Q_\pm=b^\mp f^\pm,
\eeq
which are quadratic in bosonic ($b^\pm$) and fermionic 
($f^\pm$) creation-annihilation operators.
The operator $\Gamma=[f^+,f^-]$ has a sense of the grading operator,
$\Gamma^2=1$, which classifies all the operators into
even (bosonic, $B$) and odd (fermionic, $F$) 
subsets according to the relations $[\Gamma, B]=0$, 
$\{\Gamma, F\}=0$. 
The even ($H$) and odd ($Q_\pm$) integrals of motion
form the $N=1$ superalgebra:
\beq
\label{N1}
\{Q_+,Q_-\}=H,\quad
Q_+^2=Q_-^2=0,\quad
[H,Q_\pm]=0.
\eeq
Realization of operators $f^\pm$ in terms of Pauli matrices,
$f^\pm=\sigma_\pm\equiv
\frac{1}{2}(\sigma_1\pm i\sigma_2)$,
leads to the diagonal form for the grading operator,
$\Gamma=\sigma_3$. 

Simple change in (\ref{qosc}) of bosonic operators $b^\pm$ for 
the mutually hermitian conjugate linear differential 
operators 
\beq
\label{Bpm}
B^\pm=\frac{1}{\sqrt{2}}\left(W(x)\mp\frac{d}{dx}\right)
\eeq
results in supersymmetric quantum mechanics (SUSYQM) \cite{wit,susyqm}
generalizing the simplest $N=1$ supersymmetric  system (\ref{hosc}),
(\ref{qosc}),
\beq
\label{wit}
H=\frac{1}{2}\left(-\frac{d^2}{dx^2}+W^2(x)+W'(x)\sigma_3\right),\quad
Q_\pm=B^\mp\sigma_\pm,
\eeq
with $H$ and $Q_\pm$ satisfying the same superalgebra (\ref{N1}).

The reflection operator $R$, $R^2=1$, 
$R\psi(x)=\psi(-x)$, introduces the $\ZZ_2$-grading structure
in the space of wave functions by
classifying them as
even and odd ones, $R\psi_\pm(x)=\pm\psi_\pm(x)$,
$\psi_\pm=\frac{1}{2}(\psi(x)\pm\psi(-x))$.
Since the $\ZZ_2$-grading structure is the necessary element
of SUSYQM, one can try to realize the latter without introducing
independent fermionic operators or 
associated matrix structure but by declaring the reflection operator
to be the grading operator, $\Gamma=R$.
Indeed, first we note that the operator $R$ 
gives a possibility to construct
the operators
$\Sigma_1=\epsilon(x)$, $\Sigma_2=i\epsilon(x)R$,
$\Sigma_3=R$ (with $\epsilon(x)=-1,0,+1$ for 
$x<0,=0,>0$), which satisfy the algebra of
Pauli matrices: $\Sigma_i\Sigma_j=\delta_{ij}+i\epsilon_{ijk}
\Sigma_k$. However, unlike the matrices $\sigma_i$,
the operators $\Sigma_i$ do not commute 
with $x$ and $d/dx$. In particular,
$\Sigma_3=R$ anticommutes with these
operators. 
The analogs of nilpotent operators 
$\sigma_\pm$ take here the form
$\Sigma_\pm\equiv\frac{1}{2}
(\Sigma_1\pm i\Sigma_2)=\epsilon(x)\Pi_\mp$,
where 
$\Pi_\pm=\frac{1}{2}(1\pm R)$ are the projectors,
$\Pi_\pm^2=\Pi_\pm$, $\Pi_+\Pi_-=0$, $\Pi_++\Pi_-=1$.
Such a structure of operators $\Sigma_\pm$
provides us with a hint for the
construction of odd (anticommuting with the grading operator $R$)
nilpotent supercharge operators
by analogy with (\ref{wit}), (\ref{Bpm}),
\beq
\label{qbos}
Q_\pm=\frac{1}{\sqrt{2}}\left(W_-(x) \pm \frac{d}{dx}
\right)\Pi_\pm,
\eeq
where $W_-(x)$ is a superpotential to be, unlike the case of (\ref{wit}),
an {\it odd function}. Defining the Hamiltonian
as an anticommutator of $Q_+$ and $Q_-$, we get
\beq
\label{hbos}
H=\frac{1}{2}\left(-\frac{d^2}{dx^2}+W_-^2(x)-W_-'(x)R\right).
\eeq
By the construction, it commutes with the grading operator $R$,
i.e., is the even operator,
and the operators (\ref{qbos}), (\ref{hbos})
form the $N=1$ superalgebra
(\ref{N1}). 

The formally constructed {\it minimally} bosonized supersymmetric
quantum mechanics \cite{bsusy,bsusy1} given by Eqs. (\ref{qbos}), (\ref{hbos}) 
is related to the initial version of SUSYQM (\ref{wit})
in a nontrivial way. This can be observed immediately just by
considering the simplest case of linear superpotential
$W(x)=\epsilon x$, $\epsilon=+$ or $-$.
Both these cases correspond to the same system
of superoscillator (\ref{hosc}), (\ref{qosc})
with the spectrum $E=0,1,1,2,2,\ldots$,
i.e.,  the supersymmetry in this case is exact.
On the other hand, the bosonized version
of SUSYQM with $W_-(x)=+x$ describes the 
system with exact supersymmetry ($E=0,2,2,4,4,\ldots$),
whereas $W_-(x)=-x$ gives a system in the phase of spontaneously
broken supersymmetry ($E=1,1,3,3,\ldots$) \cite{bsusy,bsusy1}.
The specified spectra of these two different bosonized
supersymmetric systems
constitute together the spectrum of
superoscillator (\ref{hosc}).
This observation helps to establish the exact relation
between the two versions of SUSYQM:
if the ordinary supersymmetric
system is given by the superpotential 
being an odd function, $W(-x)=-W(x)$,
then the bosonized version of SUSYQM can be obtained
from it by applying the special unitary transformation \cite{gpz}
\beq
\label{spu}
\psi(x)\rightarrow \tilde{\psi}(x)=U\psi(x)=
\rho_+\psi(x)-\rho_-\psi(-x),\quad
\rho_\pm=\frac{1}{2}(1\pm \sigma_1),
\eeq
with subsequent reduction to one of the eigenspaces
of $\sigma_3$.
Note that (\ref{spu}) is the 
{\it non-local} (bi-local) transformation.
Nonlocality of bosonized SUSYQM
is also hidden in the nature of the grading operator
$R$:
in coordinate representation it can be represented 
in the form  $R=\sin(\pi H_0)$,
$H_0=\frac{1}{2}(x^2-d^2/dx^2)$ \cite{bsusy1}.

Though the described relation in the direction 
(\ref{wit})$\rightarrow$(\ref{qbos}), (\ref{hbos})
is simple, the inverse correspondence is not clear:
if we are given some bosonized supersymmetric
system (e.g., with $W_-(x)=-x$),
what is the system of the form (\ref{wit}) to be
isospectral to it?

The bosonized supersymmetry can be treated 
as a supersymmetry of the system of 
two identical fermions living on a line:
it is realized in their rest frame system 
in the $j_3=0$ eigensubspace of the total vector spin operator
${\bf J}$, 
$J_i=\frac{1}{2}(\sigma_i\otimes 1+
1\otimes\sigma_i)$, $i=1,2,3$ \cite{gpz}.

Bosonized version of 
$N=1$ SUSYQM can directly be extended 
for the two-dimensional space \cite{gpz}:
\[
Q_1=\frac{1}{\sqrt{2}}(\pi_1+i\pi_2 R),\,\,
Q_2=iRQ_1,\,\,
H=\frac{1}{2}\left(\pi_a\pi_a-RB(x)\right),
\]
where $\pi_a=-i\partial/\partial x^a-A_a(x)$,
$a=1,2,$ $A_a(x)$ is an odd `electromagnetic potential',
$A_a(-x)=-A_a(x)$, $B$ is a `magnetic' field,
$B(x)=\partial_1A_2(x)-\partial_2A_1(x)$,
and $R$ is given by
$R=R_1R_2$, where $R_1$, $R_2$ are reflection
operators with respect to $x^1$ and $x^2$:
$\{R_1,x^1\}=\{R_2,x^2\}=0$,
$[R_1,x^2]=[R_2,x^1]=0$, $R_1^2=R_2^2=1$,
$[R_1,R_2]=0$. In this case the operator
$R$ has also the sense of the operator of space rotation 
by the angle $\pi$, $R=\exp(-i\pi L)$,
where $L=i(x^2\partial_1-x^1\partial_2)$
is the operator of orbital angular momentum.

There is another interesting application of the
bi-local unitary transformation (\ref{spu}):
being applied to the Dirac
field in $1+1$ dimensions, it turns out to be analogous to 
the Foldy-Wouthuysen (FW) transformation.
Indeed, let us consider the unitary transformation
$\Psi(t,x)\rightarrow \tilde{\Psi}(t,x)=U\Psi(t,x)$,
$U=\varrho_+-\varrho_- R$,
$\varrho_\pm=\frac{1}{2}(1\pm\gamma_5)$,
$\gamma_5=\gamma^0\gamma^1$,
applied to the Dirac field satisfying 
the equation 
\[
[i(\gamma^0\partial_t
+\gamma^1\partial_x)-m]\Psi(t,x)=0.
\]
Here we assume that $R$ is the space reflection (parity)
operator, $Rt=tR$, $Rx=-xR$, $R^2=1$.
In representation $\gamma^0=\sigma_3$,
$\gamma^1=i\sigma_2$,
the unitary operator $U$ takes exactly the form
of the operator defining the transformation 
(\ref{spu}).
The transformed Dirac field obeys the equation
\[
[iR(\partial_t+\partial_x)+m\sigma_3]\tilde{\Psi}(t,x)=0,
\]
or, equivalently, the equation
$i\partial_t\tilde{\Psi}=\tilde{H}\tilde{\Psi}$ with the Hamiltonian
$\tilde{H}=-i\partial_x-R\sigma_3m$,
which has a diagonal matrix form.
Though $\tilde{H}$ has not the FW square root form,
nevertheless, the relation 
$\tilde{H}{}^2=-\partial_x^2+m^2$
is valid due to the dependence of the Hamiltonian 
on the reflection operator $R$.
If, like in the SUSYQM case,
we further realize a reduction of the
system to the eigensubspaces of $\sigma_3$,
we arrive finally at the linear
differential equation with reflection
\beq
\label{sqrt}
[iR(\partial_t+\partial_x)+\epsilon m]\psi_\epsilon=0,\quad \epsilon=+,-,
\eeq
being a square root of the Klein-Gordon equation
for the one-component field $\psi_\epsilon$
since the equation $(-\partial_t^2+\partial_x^2-m^2)\psi_\epsilon=0$
appears as the consequence of (\ref{sqrt}).
The reduction, however, destroys the Poincar\'e
invariance: the theory for one-component field $\psi_\epsilon$
is not invariant under either Lorentz transformations
or space translations.

Equation (\ref{sqrt}) admits an extension for 
arbitrary dimension by introducing
the set of operators $R_i$,
$R_i x_i=-x_iR_i$ (no summation),
$R_i x_j=x_jR_i$, $i\neq j$,
$R_i t=tR_i$. E.g., in the (2+1)-dimensional
case the nontrivial operator term from Eq. (\ref{sqrt})
is substituted for $\Delta=iR_2[R_1(\partial_0+\partial_1)+\partial_2]$,
$\Delta^2=-\partial_0^2+\partial_1^2+\partial_2^2$.
The corresponding $d$-dimensional one-component square root
of the Klein-Gordon equation admits introducing
the interaction with external gauge fields (for the details 
see ref. \cite{gpz}).

It seems that the bosonized SUSYQM can be related to
the (1+1)-dimensional integrable systems on the half-line.
Besides, possibly, there is
a relation between one-component
linear differential equation (\ref{sqrt}) and
the theory of massless boson and fermion fields 
on the half-line (or relation between $d$-dimensional
generalization of Eq. (\ref{sqrt}) and the theory of corresponding
massless fields living in the subspace $x_i\geq 0$).
The latter possibility is due to the 
appearance of mass parameter in the boundary
conditions which also destroy space translation and Lorentz
symmetries \cite{lig}.

Let us return to the bosonized form of SUSYQM (\ref{qbos}), (\ref{hbos}).
In the case of the choice of the superpotential $W(x)=-\frac{\nu}{2x}$,
$\nu\in{\RR}$,
the hermitian linear combination $i(Q_--Q_+)$
of supercharges from (\ref{qbos})
takes the form of the Yang-Dunkl operator
$D_\nu=-i(\frac{d}{dx}-\frac{\nu}{2x}R)$,
related to the 2-particle Calogero model,
where $R$ plays the role of exchange operator \cite{gpz}.
With this operator, one can construct the analogs of bosonic operators
$b^\pm$, 
\beq
\label{acoor}
a^\pm=\frac{1}{\sqrt{2}}(x\mp iD_\nu).
\eeq
Together with the reflection operator $R$
they form the $R$-deformed Heisenberg algebra (RDHA) \cite{def0}--\cite{def2},
\beq
\label{RDHA}
[a^-,a^+]=1+\nu R,\quad
\{R,a^\pm\}=0,\quad
R^2=1.
\eeq
This algebra possesses infinite-dimensional
unitary representations for the values of
deformation parameter $\nu>-1$,
and at the integer values $\nu=p-1$, $p=1,2,\ldots$,
it defines the parabosons of order $p$ \cite{def,def2}.
At $\nu=-(2p+1)$, this algebra defines (deformed)
parafermions of order $p$ \cite{def,def2} (see below).

The Hamiltonian of bosonized supersymmetric 
systems with the superpotentials  
$W(x)=\epsilon x-\frac{\nu}{2x}$, $\epsilon=+,-$,
can be represented in terms of generators of algebra
(\ref{RDHA}),
\beq
\label{haa}
H_\epsilon=\frac{1}{2}\{a^+,a^-\}
-\frac{1}{2}\epsilon R[a^-,a^+].
\eeq
With the help of the number operator
\beq
\label{number}
N=\frac{1}{2}\{a^+,a^-\}-\frac{1}{2}(1+\nu),
\eeq
$[N,a^\pm]=\pm a^\pm$,
and the relation $R=(-1)^N$,
one can easily find that 
in the case $\nu>-1$, $\epsilon=+$,
the Hamiltonian (\ref{haa}) gives the family
of supersymmetric systems isospectral
to the superoscillator system (\ref{hosc})
(more exactly, to the system with the Hamiltonian
$\tilde H=2H$, where $H$ is given by (\ref{hosc})).
On the other hand, the case 
$\nu>-1$, $\epsilon=-$, describes the family of
systems being in 
the phase of spontaneously broken supersymmetry,
for which the two lowest states possess 
positive energy, whose value is governed by the
deformation parameter $\nu$: $E_0=1+\nu>0$.
When $\nu=1$, RDHA (\ref{RDHA}) describes parabosons
of order $p=2$ ($a^-a^+|0\rangle=2|0\rangle$,
$a^-|0\rangle=0$),
and the Hamiltonian (\ref{haa}) takes a normal,
$H_+=a^+a^-$, or antinormal, $H_-=a^-a^+$,
form. 

What will happen with supersymmetry
if we take the Hamiltonian in the same normal or antinormal
form in terms of $a^\pm$ operators corresponding
to the values of $\nu$ different from $1$?
Using Eqs. (\ref{RDHA}), (\ref{number}),
we find that for $\nu=2k+1$,
i.e. for parabosons of order $p=2k+2$,
the Hamiltonian $H_+=a^+a^-$ reveals
a supersymmetric structure.
The total spectrum is given by the 
series $E=0,\ldots, 2(k-1),2k,2k+2,2k+2, 2k+4,2k+4,\ldots$,
i.e., the peculiarity of such systems
consists in the presence of $k+1$ lower lying 
singlet states (with the lowest state $|0\rangle$ 
of zero energy) instead of one singlet state 
of zero energy which we have in the case of usual
supersymmetry.
Therefore, such supersymmetric
systems (for $k=1,2,\ldots$) cannot be described 
by the bosonized version of supersymmetric quantum mechanics.
To clarify their nature, it is sufficient
to find the integrals of motion
$Q_\pm$, $[H_+,Q_\pm]=0$,
being nilpotent supercharges, $Q_\pm^2=0$,
which transform mutually the supersymmetric doublet
states and annihilate singlet states.
The sought for operators have the form
\beq
\label{qnon}
Q_+=(a^+)^{2k+1}\Pi_-,\quad
Q_-=(a^-)^{2k+1}\Pi_+,
\eeq
and satisfy the relation
\beq
\label{poly}
\{Q_+,Q_-\}={\cal P}_{2k+1}(H_+),
\eeq
where ${\cal P}_{2k+1}(H_+)$ is the polynomial
in $H_+$ of order $p-1=\nu=2k+1$:
${\cal P}_{2k+1}(H_+)=(H_+-2k)(H_+-2k+2)\ldots
(H_++2k-2)(H_++2k)$.

For parabosons of order $p=2k+2$, ($\nu=2k+1$),
$k=1,\ldots$,
the systems given by the shifted antinormally ordered
quadratic Hamiltonian $H_a=H_--2$
are also characterized
by the nonlinear supersymmetry
with supercharge operators having
the form (\ref{qnon}) with projectors
$\Pi_+$ and $\Pi_-$ to be changed in their places.
In this case the polynomial appearing in the anticommutator
of supercharges is 
${\cal P}_{2k+1}(H_a)=(H_a-2k+2)\ldots (H_a+2k)(H_a+2k+2)$.
The parabosonic system with the Hamiltonian $H_a$
is characterized by the presence of $k$
singlet states with the state $|1\rangle=a^+|0\rangle$ 
playing the role of the ground state of
zero energy. 
This state is odd, $R|1\rangle=-|1\rangle$,
unlike the case of bosonized supersymmetric quantum
mechanics (\ref{qbos}), (\ref{hbos}),
where the ground state of zero
energy (if exists, i.e. in the case of exact supersymmetry)
is always even \cite{gpz}.
This explains, in particular, why 
the system described by the Hamiltonian 
$H_a$ in the case of parabosons of order $p=4$
has exactly the same spectrum as the parabosonic
system of order $p=2$ with the Hamiltonian $H_+$
(i.e. of the form of the spectrum of the 
superoscillator (\ref{hosc})), but
is characterized by the nonlinear supersymmetry with the 
polynomial ${\cal P}_3(H_a)$ of order $3$.

If we remember the coordinate
realization (\ref{acoor}),
the Hamiltonians $H_+$ and $H_a$ can be represented
in the form 
\beq
\label{hcal}
{\cal H}_\epsilon=
\frac{1}{2}\left(-\frac{d^2}{dx^2}+
x^2+\frac{\nu^2}{4x^2}-2+\epsilon
+\nu\left(\frac{1}{2x^2}-\epsilon\right)R\right),
\eeq
where $\epsilon=+1$ corresponds to $H_+$ and $\epsilon=-1$
corresponds to $H_a$.
The system given by the Hamiltonian (\ref{hcal})
can be treated as a 2-particle
Calogero-like model with exchange interaction,
where $x$ has a sense of relative coordinate
and $R$ has to be understood as an exchange operator.
Therefore, at $\nu=2k+1$ the class of Calogero-like systems
(\ref{hcal}) possesses hidden supersymmetry
with the supercharges given by
\[
Q_+=(Q_-)^\dagger=\frac{1}{2^{3(k+\frac{1}{2})}}
\left(\left(-\frac{d}{dx}+x+\epsilon
\frac{\nu}{2x}\right)(1-\epsilon R)\right)^{2k+1},
\]
and characterized by the superalgebra
of the form (\ref{poly}). In correspondence 
with the discussion above, this supersymmetry
can also be treated as a supersymmetry 
realized in the sector $j_3=0$
of the system of two identical fermions on a line.

Nonlinear (polynomial) supersymmetries appearing in the systems
of parabosons of even order can be obtained 
via appropriate modification of classical analog for
Witten's supersymmetric quantum mechanics. 
Indeed, the system described by the Lagrangian \cite{parab}
\[
L=\frac{1}{2}(\dot{x}{}^2-W^2(x)+ikW'(x)\epsilon_{ab}\theta_a\theta_b
+i\theta_a\theta_a),
\]
with $\theta_a$, $a=1,2$, real Grassmann variables,
and $\epsilon_{ab}=-\epsilon_{ba}$, $\epsilon_{12}=1$,
is characterized by the two odd integrals of motion,
\[
Q^\pm_k=(B^\pm)^k\theta^\mp,
\]
where 
$\theta^\pm=\frac{1}{\sqrt{2}}(\theta_1\pm i\theta_2),$
and 
$B^\pm=\frac{1}{\sqrt{2}}(W(x)\mp ip)$
are the classical analogs of $f^\pm$ and of operators (\ref{Bpm})
with $p$ being a momentum canonically conjugate to $x$.
Together with the Hamiltonian 
\[
H_k=\frac{1}{2}(p^2+W^2-ikW'\epsilon_{ab}\theta_a\theta_b),
\]
they satisfy the Poisson bracket relation
\beq
\label{qqhk}
\{Q^+_k,Q^-_k\}_{{}_{PB}}=-i(H_k)^k,
\eeq
which, unlike the usual case $k=1$,
is nonlinear for $k=2,\ldots$.
In the case of $W=x$, the quantization of the system
modifies the form of the algebra due to the quantum 
corrections and the quantum analog of (\ref{qqhk})
takes the form of the polynomial supersymmetry
analogous to the supersymmetry in pure parabosonic systems
with the Hamiltonians
quadratic in creation-annihilation operators.
However, in the case of the superpotential
different from the linear one,
generally the quantum anomalies arise which destroy
supersymmetry algebra \cite{parab}.
Note that the polynomial supersymmetry appeared also
in other contexts in \cite{as,is}.

The RDHA (\ref{RDHA}) related to the single-mode paraboson
systems can be given in equivalent form by
specifying it via the relations \cite{parab}
\beq
\label{aaf}
a^+a^-={\cal F}(N),\quad
a^-a^+={\cal F}(N+1),\quad
[N,a^\pm]=\pm a^\pm,
\eeq
where ${\cal F}(N)\equiv N(-1)^N+\nu\sin^2\frac{\pi N}{2}$
is the so called structure function,
satisfying for $\nu>-1$ the relations
${\cal F}(0)=0$, ${\cal F}(n)>0$, $n=1,2,\ldots$.
Then the hidden supersymmetries 
of parabosonic systems are encoded
in the symmetry relations
${\cal F}(2n+1)={\cal F}(2n+\nu+1)$ taking place
for $\nu=2k+1$, $k=0,1,\ldots$.
On the other hand, in the case $\nu=-(p+1)$, 
$p=2,4,\ldots$, the structure function
is characterized by the relation
${\cal F}(p+1)=0$, which
signals on existence of $(p+1)$-dimensional 
irreducible representations of the
algebra (\ref{RDHA}), in which the relations $(a^\pm)^{p+1}=0$
are valid. However, in this case
the function ${\cal F}(N)$ 
is not positive-definite, and the corresponding finite-dimensional
representations are not unitary.
Defining new creation-annihilation operators
$f^-=a^-R$, $f^+=a^+$, we arrive at the following parafermionic type 
algebra of the even order \cite{def}:
\[
\{f^-,f^+\}=p+1-R,\quad
(f^\pm)^{p+1}=0,\quad
\{R,f^\pm\}=0,
\]
with $p=2k$. Due to the presence of the reflection operator,
the algebra has the natural internal $\ZZ_2$ structure and
is characterized by the positive-definite structure function
\beq
\label{z2}
{\cal F}(N)=N(-1)^N+(p+1)\sin^2\frac{\pi N}{2}
\eeq
possessing the symmetry
\beq
\label{fsym}
{\cal F}(n)={\cal F}(p+1-n),\quad n=0,\ldots,p+1.
\eeq
The parafermionic type systems of order $p$
can be described by the relations of the same
form (\ref{aaf}) with structure functions
obeying the relations ${\cal F}(0)={\cal F}(p+1)=0,$
${\cal F}(n)>0$, $n=1,\ldots,p$ \cite{das,que}.
Besides the described parafermions with internal 
$\ZZ_2$ structure,
the symmetry relation (\ref{fsym}) is satisfied by the 
structure functions describing other parafermionic systems
including the ordinary parafermions
(${\cal F}(N)=N(p+1-N)$), finite-dimensional
$q$-deformed oscillator 
(${\cal F}(N)\equiv
{\cal F}_q(N)=\sin\frac{\pi N}{p+1}/\sin\frac{\pi}{p+1}$,
$q=\exp i\frac{\pi}{p+1}$) and the
$q$-deformed parafermionic oscillator
(${\cal F}(N)={\cal F}_q^2(N)$).
Due to the symmetry (\ref{fsym}), all the listed
parafermionic systems with the quadratic
hamiltonians $H_n=f^+f^-$, $H_a=f^-f^+$ and $H_s=f^+f^-+f^-f^+$
reveal a typical doubling of levels and, as a consequence,
describe the supersymmetric systems.
The details on the hidden supersymmetry in such parafermionic systems
can be found in ref. \cite{paraf}, and here we restrict ourselves
only by several general comments.
In parafermionic case, due to the symmetry relation
(\ref{fsym}), the systems with $H_n$ and $H_a$ are equivalent
and one of them can be obtained from another via
the formal substitutions $f^+\rightarrow f^-$, 
$f^-\rightarrow f^+$, $n\rightarrow p-n$.
In the case of even $p$, the spectrum of the systems 
given by $H_n$ has one singlet state ($|0\rangle$)
of zero energy, whereas for odd $p$ it contains
additional singlet level ($|(p+1)/2\rangle$)
of non-zero energy. As a consequence,
in the first case the corresponding anticommutator
of supercharges is linear in Hamiltonian,
whereas in the second case it is given by the
quadratic in $H_n$ polynomial.
The Hamiltonian $H_s$ describes supersymmetric
systems in the phase of spontaneously broken 
supersymmetry (there are no singlet states of zero energy),
in which all the states are supersymmetric doublets
for odd $p$ and there is one singlet state ($|p/2\rangle$)
of positive energy for even $p$. In both cases the supersymmetry
is linear.
In parafermionic systems with quadratic Hamiltonians
more complicated cases
characterized by the anticommutator of supercharges
being the polynomial of order higher than $2$ 
can also be obtained. 
E.g., in the case of ordinary parafermions (${\cal F}(N)=N(p+1-N)$)
of order $p\geq|k|+1$, $|k|\geq 2$,
the systems given by the Hamiltonian
\[
H_k=\{f^+,f^-\}+k[f^+,f^-]+p(|k|-1), \quad k\in \ZZ,
\]
are characterized by the polynomial supersymmetry
whose order is $|k|$ ($|k|+1$) when the difference $p-k$ is odd (even).

We conclude that the hidden supersymmetries of pure parabosonic
and parafermionic systems can be explained on the general 
ground by the corresponding symmetry properties 
of the structure functions. At the same time, 
the possibility of realizing supersymmetry
in pure parafermionic case is not so surprising:
parafermions generalize fermions
with their basic relations $\{f^+,f^-\}=1$, $(f^\pm)^2=0$,
which can be treated as specifying a trivial supersymmetric
system with the supercharges $Q_\pm=f^\pm$ and the Hamiltonian
$H=1$.
On the other hand, the existence of hidden supersymmetries
in pure parabosonic systems with quadratic Hamiltonians
seems to be rather peculiar.
It can be given the following interpretation.
One can
represent the corresponding supercharge operators
(\ref{qnon}) in terms of only creation-annihilation
operators by using the
relation $R=(-1)^N$ and Eq. (\ref{number}).
In this way we write the projectors $\Pi_\pm$
in the equivalent form \cite{parab}
\[
\Pi_+=\cos^2{\cal N},\quad
\Pi_-=\sin^2{\cal N},\quad
{\cal N}=\frac{\pi}{4}\{a^+,a^-\}.
\]
Therefore, though the corresponding 
Hamiltonians $H_+=a^+a^-$ and $H_a=a^-a^+-2$
are quadratic in operators $a^\pm$,
the price we are paying to have supersymmetry
in pure parabosonic systems is the 
specific structure of the supercharges: they are
the infinite series in $a^\pm$ of the special form
which gives rise to their nilpotency and to
the polynomial form for the anticommutator (\ref{poly}).
However, it is not clear what are the corresponding supersymmetry
transformations for such pure parabosonic systems
and for the related minimally bosonized supersymmetric 
quantum mechanics. Probably, the construction of
classical models for them could help to answer this question.

\paragraph*{Acknowledgements}
The work was supported in part  by 
the grant 1980619 from FONDECYT (Chile)
and by DICYT (USACH).

\end{document}